\documentstyle[twocolumn,prb,aps]{revtex}

\begin{document}
\title{Density modulation and electrostatic self-consistency in a 
two-dimensional electron gas subject to a periodic quantizing 
magnetic field}

\author{Ulrich J. Gossmann, Andrei Manolescu$^*$ and Rolf R. Gerhardts}
\address{
Max-Planck-Institut f\"ur Festk\"orperforschung, Heisenbergstrasse 1,
D-70569 Stuttgart, Federal Republic of Germany\\
$^*$Institutul de Fizica \c{s}i Tehnologia Materialelor, C.P. MG-7
Bucure\c{s}ti-M\u{a}gurele, Rom\^ania}

\maketitle

\begin{abstract}
We calculate the single-particle states of a two-dimensional electron gas 
(2DEG) in a perpendicular quantizing magnetic field, which is periodic in 
one direction of the electron layer. We discuss the modulation of the 
electron density in this system and compare it with that of a 2DEG in a 
periodic electrostatic potential. We take account of the induced potential 
within the Hartree approximation, and calculate self-consistently the 
density fluctuations and effective energy bands. The electrostatic effects 
on the spectrum depend strongly on the temperature and on the ratio 
between the cyclotron radius $R_c$ and the 
length scale $a_{\scriptscriptstyle\!\delta\!\rho}$ of the density 
variations. We find that $a_{\scriptscriptstyle\!\delta\!\rho}$ 
can be equal to the modulation period $a\,$, but also much smaller. 
For $R_c \sim a_{\scriptscriptstyle\!\delta\!\rho}$ the spectrum in the 
vicinity of the chemical potential 
remains essentially the same as in the noninteracting system, 
while for $R_c \ll a_{\scriptscriptstyle\!\delta\!\rho}$ it may be drastically 
changed by the Hartree potential: For noninteger filling factors the energy 
dispersion is reduced, like in the case of an electrostatic modulation, 
whereas for even-integer filling factors, on the contrary, the dispersion 
may be amplified.
\end{abstract}


\section{Introduction}
\label{secin}
The interest in nonuniform magnetic fields, with spatial variations on a
nanometer scale, has been stimulated by several recent experimental 
realizations, like magnetic quantum wells \cite{McAw90} or magnetic 
superlattices.~\cite{CaEA95,Ye95a,IzEA95,Ye95b}
In the quasi-classical regime of low 
magnetic fields, the theoretical investigations have 
concentrated on the commensurability oscillations of the 
resistivity,~\cite{VaPe90,XuXi92,PeVa93,Ger96a} which are equivalent 
to the Weiss oscillations \cite{WOS88,WOS89} that occur in the 
presence of a periodic electrostatic potential. 

The quantum regime of nonuniform magnetic fields with a strong 
variation of the order of 1 T within a distance of a few hundred 
nanometers is now experimentally accessible.~\cite{WY} 
For this regime single-particle quantum mechanical calculations,
concerning the tunneling through magnetic barriers or the bound states 
in magnetic wells, have recently been performed by Peeters, Matulis, and 
Vasilopoulos.~\cite{PeMa93,MaPeVa94}   
Coulomb interaction effects have been discussed by Wu and Ulloa
\cite{WuUl93} who studied the electron density distribution and the 
collective excitations in a magnetic superlattice with a short period, 
comparable to the average magnetic length. 
They found that the periodic magnetic field
gives rise to an electron-density modulation, which is reduced due to the
counteracting induced electric field.  

In the present paper we consider a two-dimensional electron gas (2DEG) in a
perpendicular magnetic field of the form $B = B_0 + B_{mod}$ where $B_0$ is
a homogeneous part and $B_{mod}$ is, in the plane of the 2DEG, 
periodic in one direction with zero average. We describe in detail the 
charge-density response to the periodic part and  the effects of the 
associated electrostatic potential. It will be very instructive to compare 
this situation with the modulation by a unidirectional periodic 
electrostatic potential $V_{ext}(x)$ (we include the charge $-e$ in 
the definition of the electrostatic potentials, which are therefore 
rather potential energies). 
The homogeneous part of the magnetic field is 
assumed to be strong enough so that a description in terms of 
Landau levels is adequate; we denote by
$l_0=\sqrt{\hbar/eB_0}$ and by $\omega_0=eB_0/m$ the magnetic length and 
the cyclotron frequency associated with the uniform field $B_0$.
For a fixed mean electron density $\rho_0$, 
the number of relevant Landau levels is of the order of the filling
factor  $\nu = (2\pi l_0^2) \rho_0$  and thus inversely proportional to the
average magnetic field.

We first note that classically a magnetic field - modulated or not - 
has no influence in thermodynamic equilibrium because it drops out of 
the integral over momenta in the partition function
(Bohr-van Leeuwen theorem \cite{Vleck}). Especially a magnetic modulation does
not lead to a position dependence of the electron energy, which remains 
$mv^2/2$, and does not affect the equilibrium electron density. 
In contrast one expects (e.~g.~from
Thomas-Fermi-theory) that modulation by an electrostatic potential should  
lead also to a modulation of the density 
$\delta\rho(x) = -(V_{ext}(x)/\mu)\rho_0$,  where $\mu$ is the 
chemical potential. In the classical limit, i.~e.~for 
low average magnetic field with $\hbar\omega_0 \ll k_BT$,  
the density response of the 
2DEG to electrostatic and to magnetic modulations is thus very
different. A pure magnetic modulation does classically not 
give rise to electrostatic effects, whereas an external 
electrostatic potential is screened by the induced Hartree potential.

However, in the quantum regime of low filling factors the two types of
modulation affect the density in a very similar manner: Both lift the
degeneracy of the Landau levels and lead to dispersive bands. 
The homogeneous part $B_0$ of the magnetic field restricts the spatial 
extent of the relevant wavefunctions to the order of the 
cyclotron radius $R_c = l_0\sqrt{2n_{F}+1}$ where $n_{F}$ is the 
index of the Landau level at the chemical potential 
(that is $n_{F}$ is the largest integer smaller than $\nu/g_s$; 
we assume spin degeneracy $g_s = 2$ in this work). 
If $n_{F}$ is small the modulation is typically slowly varying on the 
scale $R_c$ and we can represent the Landau levels as functions 
$\epsilon_n(x)$, varying on the same length scale as the modulation. 
The width $\Delta\epsilon_n$  of these bands is of the order of 
$\hbar e\Delta B/m$ or $\Delta V$, where $\Delta B$ and $\Delta V$ 
stand for the amplitude of the modulation of magnetic field and 
electrostatic potential, respectively. In this situation we expect 
that the density is determined by a ``local'' filling  factor 
$\nu(x) = g_s \sum\limits_n f(\epsilon_n(x))$, where $f(\epsilon)$ 
denotes the Fermi function. For temperatures satisfying 
$k_BT \ll \Delta\epsilon_n\,$,  
this must lead to a density modulation 
of order $(g_s/2\pi l_0^2)$ for both electric and magnetic modulations. 

For an electric modulation it is known that, due to this
strong effect of the Landau level dispersion on the density, 
the inclusion of the Coulomb interaction in the
Hartree-approximation (which we will refer to as the electrostatically
self-consistent system) changes drastically the spectrum of the
system for low filling factors. 
Wulf {\sl et.~al.} \cite{WuGuGer88} found that, for filling factors 
not too close to an even-integer value, 
the self-consistent result corresponds to a nearly
perfect screening of the modulating external potential and the Landau bands
are flat within the order of $k_BT$. If for even-integer 
filling factor the chemical potential lies in a gap, 
the screening is much weaker although still
considerable. For potentials strong enough to yield overlapping bands, 
$|\Delta\epsilon_n| > \hbar\omega_0\,$, the screening 
around even-integer filling factors becomes nonlinear, 
featuring two bands touching the Fermi level, and the width
of the band $n_{F}$ therefore locked to $\hbar\omega_0$.

The width of the Landau band at the Fermi level, however, 
is the basic element for understanding transport measurements
\cite{VaPe90,Ye95a,ZhGer90} and a major
aim of this work is to investigate its behaviour for a magnetic modulation 
when electrostatic self-consistency is properly accounted for. 

The paper is organized as follows. In section \ref{secdm} we describe our
model and the self-consistency problem in detail. In section \ref{seclp}
we treat the case of low filling factors, corresponding to a 
magnetic modulation varying slowly on the length scale $R_c$. 
In section \ref{secms} we discuss the regime of lower average magnetic 
fields, where the cyclotron radius is not small against 
the period of the modulation. The numerical
results we present are obtained using the material parameters of GaAs, 
namely the effective mass $m = 0.067m_e$ and 
the dielectric constant $\kappa=12.4$. The average electron density 
is fixed to $\rho_0 = 2.4\cdot 10^{11}\mbox{ cm}^{-2}\,$, 
chosen such that $\nu B_0 = 10 \mbox{ T}$, the
period of the modulation is $a = 800 \mbox{ nm}$ and we consider values of
$B_0$ between $10 \mbox{ T}$ and $0.1 \mbox{ T}$ corresponding to cyclotron
radii between $7 \mbox{ nm}$ and $740 \mbox{ nm}$.

\section{Description of the Model}
\label{secdm}
We consider an idealized 2DEG confined to the plane \{{\bf r}=(x,y)\} and
subject to a magnetic field ${\bf B}({\bf r})=(0,0,B(x))$ 
which,~\cite{XuXi92,PeVa93,Ger96a,WuUl93}  
within the plane, is directed in $z$-direction, does not depend
\cite{GerPfGu96} on $y$, and has a simple
periodic dependence on $x\,$:
\begin{equation}
B(x)=B_0+B_1\cos Kx \, ,
\label{mmod}
\end{equation}
where $K=2\pi/a$ is the wave vector of the modulation.
We start with the noninteracting 2DEG described by the 
standard single-electron Hamiltonian
$H^0=({\bf p} + e{\bf A})^2/2m$ in which we use the Landau gauge
for the vector potential, ${\bf A}(x)=(0,B_0x+(B_1/K)\sin Kx,0)$.
The eigenfunctions of $H^0$ depend on $y$ only through a 
plane-wave prefactor,
\begin{equation}
\psi_{nX_0}(x,y)=L_y^{-1/2}e^{-iX_0 y/l_0^2}\phi_{nX_0 }(x),
\end{equation}
with $L_y$ being a normalization length and $X_0$ the center coordinate.  
The functions $\phi_{nX_0}(x)$ are the  eigenvectors of the 
one-dimensional Hamiltonian
\begin{equation}
H^0(X_0)=\hbar\omega_0\left[-\frac{l_0^2}{2}\frac{d^2}{dx^2}+
\frac{1}{2l_0^2}\left(x-X_0+\frac{s}{K}\sin Kx\right)^2\right] \, ,
\label{reha}
\end{equation}
where $s=B_1/B_0$ will be refered to as the modulation strength.

For the homogeneous system, $s=0$, the functions $\phi_{nX_0}(x)$
are oscillator wave functions centered on $X_0$,
also known as Landau wave functions $\varphi^L_{nX_0}$, 
associated with the degenerate Landau levels 
$\varepsilon^L_n=(n+1/2)\hbar\omega_0$.  The degeneracy is lifted 
for $s\neq 0$, and the resulting energy bands $\epsilon_{n}(X_0)$, 
together with the corresponding wave functions, 
can be obtained by diagonalizing the 
reduced Hamiltonian (\ref{reha}) in the basis of the Landau wave 
functions.  The matrix elements can be written as 
\begin{eqnarray}
\label{mel}
&&\langle\varphi^L_{nX_0} \mid H^0(X_0)\mid\varphi^L_{n'X_0} \rangle \;=\; 
\nonumber \\ &&
\hbar\omega_0\Bigg\{\,\left(n+\frac{1}{2}\right)\delta_{nn'}
+\frac{s}{2z}\left[E_{nn'}(z)+\sqrt{nn'}E_{n-1,n'-1}(z) \right.
\nonumber \\ &&
\left. -\sqrt{(n\!+\!1)(n'\!+\!1)}E_{n+1,n'+1}(z)\right]
\cos\left(KX_0+(n\!-\!n')\frac{\pi}{2}\right)  
\nonumber \\ &&
+\frac{s^2}{8z}\left[\delta_{nn'}-E_{nn'}(4z)
\cos\left(2KX_0+(n\!-\!n')\frac{\pi}{2}\right)\right]\,\Bigg\}\,,  
\end{eqnarray}
where $z=(Kl_0)^2/2$. We have used the notation:
\begin{eqnarray}
\label{efn}
E_{nn'}(z) & \;=\;&
\left(\frac{n'!}{n!}
\right)^{1/2}e^{-z/2}z^{(n-n')/2}L^{n-n'}_{n'}(z) 
\nonumber \\ 
& \; = \; & (-1)^{n-n'}E_{n'n}(z) \quad,
\end{eqnarray}
with $L^m_n(z)$ being a Laguerre polynomial. Applying first-order 
perturbation theory we get from (\ref{mel}) the energy levels as 
simple cosine-shaped bands
\begin{equation}
  \label{enlpt1}
\epsilon^{PT1}_n(X_0) = \hbar\omega_0 \Big[(n\,+\,1/2\,)\,+
 s\,G_n(z)\,\cos(KX_0)\Big]
\end{equation}
where the factor $2s\hbar\omega_0G_n(z)$ $ = $ 
$s\hbar\omega_0\,e^{-z/2}(2L^1_n(z) - L^0_n(z))$ 
has an oscillatory dependence on the ratio $l_0\sqrt{2n+1}/a$ which is the
basic reason for the commensurability oscillations seen in transport
experiments.~\cite{Ye95a,PeVa93}  
The limits of validity of Eq.~(\ref{enlpt1}) will be discussed below.

The single-particle density is given by the formula
\begin{equation}
  \label{dens}
  \rho(x) = \frac{g_s}{2\pi l_0^2}\sum\limits_{n=0}^{\infty}
\int\limits_{-\infty}^{+\infty} dX_0\,f(\epsilon_n(X_0))
\mid\phi_{nX_0}(x)\mid^2
\end{equation}
where $f(\epsilon)$ denotes the Fermi function and $g_s=2$ 
accounts for spin degeneracy. 

The density determines the electrostatic (Hartree) potential, 
which we treat by Fourier expansion  
$V^H(x) = \sum\limits_{\eta \ge 1} V^H_{\eta} \cos(\eta K x)$. Here
\begin{equation}
  \label{vhnu}
  V^H_{\eta} = \frac{e^2}{4\pi\epsilon_0\kappa}\,
\frac{a}{\eta}\,\rho_{\eta}
\; =\;  \frac{1}{\eta} \;\;\frac{a}{2\pi\,a_B} 
\;\; (2\pi l_0^2 \,\rho_\eta)\;\; \hbar\omega_0 \;,
\end{equation}
with  $\rho(x) = \sum\limits_{\eta \ge 0}\rho_{\eta} \cos(\eta K x)$ and 
$a_B$ the effective Bohr radius. For GaAs, 
$2\pi\,a_B  \approx 63 \mbox{ nm}$. We assume that the system is 
electrically neutral such that the average density $\rho_0$ does not 
contribute to $V^H$ but only determines the chemical potential 
contained in the Fermi function. The Hartree potential has to be added 
to the Hamiltonian (\ref{reha}) and gives a contribution 
\begin{eqnarray}
    \label{melv}
&& \langle\varphi^L_{nX_0} \mid V^H(x)\mid\varphi^L_{n'X_0} \rangle
\nonumber \\ &&
\;\;=\;\sum_\eta V^H_{\eta} E_{nn'}(\eta^2z)
\cos\left(\eta K X_0+(n-n')\frac{\pi}{2}\right)
\end{eqnarray}
to the matrices (\ref{mel}). The strongest influence on the induced 
potential originates from the low Fourier components of the density, 
which are related to the long-range charge fluctuations. 
We diagonalize the Hamiltonian $H^0(X_0) + V^H$ self-consistently with 
Eq.~(\ref{dens}) by a numerical iterative scheme.
 
To understand the way the system achieves self-consistency, we occasionally
consider also a 2DEG subject to a homogeneous magnetic field $B_0$ and
a cosine electrostatic potential
\begin{equation}
  \label{vmod}
V_1(x) = V_1\,\cos(Kx)  
\end{equation}
with modulation strength $v_1 = V_1/\hbar\omega_0$ instead of the
magnetically modulated system (\ref{mmod}). First-order perturbation theory
yields for the electric modulation (\ref{vmod}) the spectrum
\begin{equation}
  \label{enelpt}
  \epsilon^{PT1}_n(X_0) = \hbar\omega_0 \Big[(n\,+\,1/2\,)\,
  +\,v_1\,F_n(z)\,\cos(KX_0)\Big]
\end{equation}
with $F_n(z) = e^{-z/2}L_n(z)$. 
\section{The Limit of Long Period}
\label{seclp}
In this section we deal with a long-period magnetic modulation, with a
strong average magnetic field, such that $Kl_0\ll 1$ and $R_c \sim l_0$ 
(but not necessarily with $B_1\ll B_0$). Approximate analytical results for
both electric and magnetic modulations will
be developed for a better understanding of the energy spectra and electron
density. We first describe the properties of the noninteracting system.

\subsection{Noninteracting electrons}
\label{seclpni}
Treating the magnetic modulation as a perturbation 
presents some difficulties in 
this limit, since for nonvanishing $s$ and $z\rightarrow 0$ 
the matrix elements 
$\langle\varphi^L_{n\,X_0}\mid H^0(X_0) \mid\varphi^L_{n+m\, X_0}\rangle$
given by Eq.~(\ref{mel}) diverge 
for $m=0,\pm 1$, while those with $m=\pm 2$ are finite and 
those with $\mid\! m\!\mid\; > 2$ vanish.  Thus the Hamiltonian matrix 
becomes band-diagonal, and the divergent elements cancel in 
a complicated way in order to yield finite eigenvalues. 
Therefore, for $Kl_0\ll 1$ an accurate numerical diagonalization 
requires a large matrix (\ref{mel}) and the Landau level 
mixing is strong, except if $s\rightarrow 0$. This complication 
does not occur for the electric modulation (\ref{vmod}) for which  
the Landau wave functions diagonalize the matrix in the long-period limit  
for any $v_1$, and first-order perturbation theory gives the 
{\em exact} energy spectrum for $z\to0$, namely~\cite{ZhGer90} 
$\epsilon_{nX_0}=(n+1/2)\hbar\omega_0+V\cos( KX_0)\,$.

Instead of using standard perturbation theory with respect to the modulation
stregth $s$, we can handle the Hamiltonian 
(\ref{reha}) by performing a Taylor expansion of the potential term 
$\Big(\hbar\omega_0/2l_0^2\Big) \left(x-X_0+\frac{s}{K}\sin Kx\right)^2$ 
around its minimum $X_1$ given by
\begin{equation}
X_1=X_0-\frac{s}{K}\sin KX_1\,. 
\label{scc}
\end{equation}
For fixed $KX_0$ and $| s | < 1$ this has a unique solution $KX_1$ with
$X_1=X_0$ for $KX_0 = 0,\pi$.  The parabolic approximation reads 
\begin{equation}
H^0(X_0)\approx\hbar\omega_0\left[-\frac{l_0^2}{2}\frac{d^2}{dx^2}+
\frac{(1+s\cos KX_1)^2}{2l_0^2}(x-X_1)^2\right] \,,
\label{apreh}
\end{equation}
with an error term of order $ s\,\hbar\omega_0 Kl_0 ((x-X_1)/l_0)^3 $ from
which we can show that (\ref{apreh}) yields the eigenvalues and the density
for low filling factors correct to leading order in $Kl_0$. 
The Hamiltonian (\ref{apreh}) is equivalent to the unperturbed one,
Eq.(\ref{reha}), but with modified center coordinate $X_1$, 
cyclotron frequency $\tilde\omega_0=\omega_0(1+s\cos(KX_1))$ 
and magnetic length $\tilde l_0=l_0/\sqrt{1+s\cos(KX_1)}$. 
The main effect of the magnetic modulation on the wave functions is 
the shift (\ref{scc}) of their center of weight. We see that, since 
$K(X_1-X_0)$ is independent of $K$, the absolute shift $X_1-X_0$ 
increases with increasing modulation period ($K\to0$) at equivalent positions
within the period (i.~e.~for fixed $KX_0$)
except for $KX_0 = 0,\pi$. This
explains the difficulties with the standard perturbation theory which expands
the shifted Landau functions with center $X_1(X_0)$ in the basis of Landau
functions centered around $X_0$.

The Landau bands resulting from Eq.(\ref{apreh}) are
\begin{equation}
\epsilon_{n}(X_0)=\hbar\omega_0 \Big(1+s\cos\big(KX_1(X_0)\big)\Big)
\left(n+\frac{1}{2}\right)\,.
\label{aplb}
\end{equation}
The appearance of $X_1$ 
instead of $X_0$ in Eq.~(\ref{aplb}) leads to a substantial 
deviation of the simple cosine band shape predicted by 
first-order perturbation theory; the bandwidth is, however, 
given correctly by Eq.~(\ref{enlpt1}). In calculating the density,  
the indicated replacement of $l_0$ by $\tilde l_0$ leads to 
corrections of higher order in $Kl_0$ and, since this order is not included
correctly, is not to be used. 
We therefore insert just shifted Landau-functions into 
Eq.~(\ref{dens}) and obtain
\begin{eqnarray}
  \label{densx1}
&& \rho(x)  = \;
 \frac{g_s}{2\pi l_0^2}\sum\limits_{n}\int\limits
dX_0\,f(\epsilon_n(X_0))\mid\varphi^L_{n,X_1(X_0)}(x)\mid^2
\nonumber \\ && 
  = \;\frac{g_s}{2\pi l_0^2}\sum\limits_{n}\int\limits
dX_1\frac{dX_0}{dX_1}\,f(\epsilon_n(X_0(X_1))
\mid\varphi^L_{n,X_1}(x)\mid^2 .
\end{eqnarray}
From (\ref{scc}) we have $dX_0/dX_1 = 1 + s\cos(KX_1)$ and with 
(\ref{aplb}) for the energy spectrum we finally get
\begin{eqnarray}
  \label{denexp}
    && \rho(x) = 
 \frac{g_s}{2\pi l_0^2}\sum\limits_{n}\int\limits
  dX_1\Big(1 + s\cos(KX_1)\Big)\,
\nonumber \\ && \times\,
 f\Big(\hbar\omega_0 \big[1+s\cos(KX_1)\big]\big(n+1/2\big)\Big)
 \mid\varphi^L_{n,X_1}(x)\mid^2 .
\end{eqnarray}
Both results (\ref{aplb}) and (\ref{denexp}) turn out to be reliable 
within a relative  accuracy of $(s\,KR_c)$ . 
They can also be derived by a simple
variational approach, using a set of translated oscillator states
$\varphi^L_{n,X_0+u}(x)$ as trial wave functions and taking the limit 
$z\to0$ after minimizing the expectation value of the 
energy.~\cite{Man92a} The numerical results which we shall present 
are obtained from a diagonalization of (\ref{mel}), however.

We note that for the electric modulation (\ref{vmod}) the results 
corresponding to Eqs.~(\ref{scc}), (\ref{aplb}) and (\ref{denexp})  read 
$X_1=X_0\;+\;v_1Kl_0^2\sin KX_1$,
\begin{eqnarray}
  \label{venx}
&& \epsilon_{n}(X_0)=\hbar\omega_0 \Big[n+1/2\; 
\nonumber \\ &&
\;\;+\;v_1\Big(1 - (1/2)\big(n+1/2\big)(Kl_0)^2\Big)\cos(KX_1)\Big]
\end{eqnarray}
and
\begin{eqnarray}
  \label{vdens}
&& \rho(x) = 
 \frac{g_s}{2\pi l_0^2}\sum\limits_{n}\int\limits
  dX_1\Big(1 - v_1(Kl_0)^2\cos(KX_1)\Big)\,
\nonumber \\ && \times \,
 f\Big(\hbar\omega_0 \big[n+1/2 +
 v_1\cos(KX_1)\big]\Big)\mid\varphi^L_{n,X_1}(x)\mid^2 . 
\end{eqnarray}
The error term in the Taylor expansion around $X_1$ is here of order $
v_1\,\hbar\omega_0 (Kl_0)^3 ((x-X_1)/l_0)^3$ and permits inclusion of the
$(Kl_0)^2$ terms. The total width of the bands from (\ref{venx}) is also
obtained from the result (\ref{enelpt}) of perturbation theory by expansion
around $Kl_0 =0$ up to order $(Kl_0)^2$. 

We see from Eqs.~(\ref{venx},\ref{vdens}) that for a long-period 
cosine {\em electric} modulation 
the bands follow the potential with constant width and the 
states are not changed by the modulation; consequently the 
density is only affected by the dispersion of the levels via the 
argument of the Fermi function. 
In contrast, for a {\em magnetic} modulation 
according to (\ref{aplb}) 
the widths of the Landau bands increase linearly with $n$ and the
$X_1$-dependent prefactor of the Fermi function in (\ref{denexp}) 
does not decrease with increasing period.

In Fig.~\ref{fig1} the dashed lines show the modulation 
of the density of the noninteracting system with a magnetic modulation of
amplitude $B_1=0.1\mbox{ T}$ for different values of the filling factor 
between 4 and 6
obtained by sweeping $B_0\,$; the
temperature is 1 K, so that $k_BT$ is much smaller than $s\hbar\omega_0$.
The density is given in units of $1/(2\pi l_0^2)$ so that the mean value of
each line equals $\nu$. 
The lines for the even-integer values of the filling factor, $\nu =4,6$,
 are marked with circles. They show a cosine form, where the amplitude is
 larger for $\nu=6$ than for $\nu =4$. 
This behaviour is easily derived from Eq.~(\ref{denexp}); 
since here the chemical potential lies in a gap, the Fermi function 
is either $0$ or $1$, and the integral gives to leading order in $Kl_0$ 
\begin{equation}
  \label{dennev}
  \delta\rho(x)\mid_{\nu\;\mbox{\tiny small and even}\,} 
= (s\nu/2\pi l_0^2) \cos Kx \quad.
\end{equation}
In the corresponding result for the electric modulation, the
factor  $s$ is replaced by
$-v_1(Kl_0)^2$ which has a different sign and vanishes for $Kl_0\to0$. 
The persistence of an finite density modulation at even-integer 
filling factors for a period much longer than the magnetic length 
constitutes a major difference between the two 
types of modulation for strong average magnetic fields. 
Note that the result (\ref{dennev}) can also be written in the form 
$\rho(x) = \nu/2\pi l^2(x)$ ($\nu$ small and even)
where $l(x) = \sqrt{\hbar/eB(x)}$ is the magnetic length 
corresponding to the local field  $B(x)$. 
This means that we can in the long-period limit think of
the  magnetic modulation as changing the local degeneracy of the 
Landau levels, thus leading to a modulated density even for spatially 
constant filling factor (in this work we use the notion of an 
$x$-dependent filling factor $\nu(x)$ as just counting the number of 
locally occupied bands, which makes sense of course only in the 
long-period limit). 
  
The dashed lines between the ones
with circles in Fig.~\ref{fig1} show the behaviour of the density while the
$n=2$ level is successively filled. Due to the energy dispersion
(\ref{aplb}) and the low temperature, the $n=2$ states around 
$KX_0 = \pi$ are occupied first, forming a region with local 
filling factor $\nu(x) = 6$ while
around $KX_0 = 0,2\pi$ we still have $\nu(x) = 4$ until the total filling
closely approaches $6$. Since the spatial extent of the wavefunctions 
is small compared to the period $a$, 
the difference in density between these two regions is of
order $g_s/(2\pi l_0^2)$. We observe, however, that within a region of
constant local filling factor the density is not constant but follows the
cosine shape imposed by Eq.(\ref{dennev}) with $\nu$ replaced by the
appropriate local filling factor $\nu(x)$.

\subsection{Self-Consistent System}
\label{seclpsc}
Since the density profiles of the noninteracting system 
correspond, according to Eq.~(\ref{vhnu}), to electrostatic
potentials with amplitudes larger than $\hbar\omega_0$, we expect 
substantial changes in the spectrum when we take 
electrostatic self-consistency properly into account, as we do now. 
The resulting densities are plotted
as solid lines in Fig.~\ref{fig1} and show much smaller fluctuations. In 
Fig.~\ref{fig2} we show results for a magnetic modulation
of $B_1 = 0.1 \mbox{ T}$ at filling factors (a) $\nu = 5$, 
(b) $\nu = 4$ and (c) $\nu = 14.3$ corresponding to average fields 
$B_0 = 2.0 \mbox{ T}, 2.5\mbox{ T}$ and $ 0.7\mbox{ T}$, respectively. 
The upper panel displays the
self-consistent spectra for temperatures $T = 1\mbox{ K}$
and $T= 0.1 \mbox{ K}$ together with the noninteracting spectra, 
the lower panel shows the corresponding self-consistent densities. 
More data for the self-consistent bandwidths and the density amplitudes in this
regime are also displayed in Fig.~\ref{fig5} and \ref{fig6} 
which are discussed in section \ref{secms-sc}. 

When the total filling factor is small and not too close to an even-integer
 value, the 
regions of increased density correspond to the minima of the Landau bands 
(cf. Fig.~\ref{fig1}). Therefore the generated Hartree potential, 
which is maximum at maximum  electron density, will act to 
{\em reduce} the dispersion of the not fully occupied band with index $n_F$. 
The self-consistent solution yields then a very flat
(``pinned'') band with deviations of only the order of $k_BT$ from 
the chemical potential, and the local filling factor 
is fractional over the whole period. This
situation is shown in Fig.~\ref{fig2} (a) for an odd-integer average 
filling factor $\nu=5$. 
The self-consistent potential here has to cancel 
the dispersion of the not fully
occupied level $n_{F}$, which is larger than the dispersion of the levels
with $n < n_F$. Thus the potential is   
$V^H(x) \approx - s\,\hbar\omega_0\,(n_{F} + 1/2)\cos Kx$ and the 
dispersion of the levels with $n < n_{F}$ is reversed in sign. 

For even-integer filling factor, however, according to Eq.~(\ref{dennev}) 
the regions of increased density 
correspond to maxima of the Landau bands, since both $\rho(x)$ and 
$\epsilon_n(X_0)$ follow the shape of the magnetic modulation with 
positive sign. Consequently, the potential generated by the 
density modulation (\ref{dennev}) {\em increases} the dispersion of the 
highest occupied band $n_{F}$ instead of acting against the 
modulation broadening. If the magnetic modulation is sufficiently weak, 
the resulting  self-consistent potential can be calculated by combining 
Eqs.~(\ref{denexp}), (\ref{vdens}) and (\ref{vhnu}) as 
$V^H(x) = \tilde{V}_H \cos Kx$ ($\nu$ small and even)
where
\begin{equation}
  \label{pinkrit}
 \tilde{V}_H = \frac{sw\hbar\omega_0\nu}{1+w(Kl_0)^2\nu}
\end{equation}
and $w = a/2\pi a_B \gg 1$. This linear behaviour breaks down, 
however, if the resulting bandwidth 
$|\Delta\epsilon_{n_F}| = 2s\hbar\omega_0(n_F+1/2)
 + 2\tilde{V}_H$ exceeds $\hbar\omega_0$.
In this case the next-higher band reaches the chemical potential 
around $KX_0 = \pi$ and
the self-consistent solution (shown in Fig.\ref{fig2} (b) for $\nu = 4$)
features a region around $KX_0 = 0,2\pi$ where the band $n_{F}$ is 
pinned to $\mu$, a region around  $KX_0 = \pi$
where the band $n_{F}+1$ is pinned to $\mu$ and a region in 
between where the chemical potential lies in a gap and the 
density still follows the cosine shape (\ref{dennev}).  
As described in section \ref{secin}, similar effects of 
electrostatic self-consistency are obtained for an electrically modulated 
system,~\cite{WuGuGer88} but there the bandwidth of the highest occupied 
band is always reduced compared to the noninteracting results. 
Then, a modulation strength $v_1>1$ 
is needed to produce the formation of pinned regions at even-integer 
filling factors, 
whereas in the magnetic case only $sw\sim 1$ must be satisfied.

For the parameters of Fig.~\ref{fig2} (c) 
the bands of the noninteracting system do overlap at the Fermi level due to
the linear increase of their width with $n$. 
In this situation the density fluctuations 
and the induced potentials consist mainly of higher
Fourier components. The corresponding wavelengths 
$2\pi/\eta K$, with $\eta>1$, are comparable to $R_c$, 
even though $R_c \ll a$ is still satisfied. 
We therefore cannot discuss the effects of the 
Hartree potential here within the limit of a long period but instead we 
have to consider the density response for electric modulations with 
$a\sim R_c$. This is done in the next section. We observe, however, 
that in Fig.~\ref{fig2} (c) the spectrum around the Fermi level 
remains essentially unchanged although we can
tell from the behaviour of the lowest level that a considerable 
electrostatic potential does exist.

For the lower temperature $T = 0.1\mbox{ K}$ the density traces in  
Fig.~\ref{fig2} (b) and very pronounced in (c) show also superimposed 
short-period oscillations. These have their origin in the nodes and 
maxima of the wavefunctions and can also be reproduced by 
Eq.~(\ref{denexp}) with the Landau functions $\varphi^L_{n}(x)$.

\section{$R_c$ comparable with Period}
\label{secms}
In this section we discuss properties of the modulated system obtained when
for fixed modulation amplitude $B_1$ the average field $B_0$ is lowered
so that we enter the regime where $R_c$ is no longer small compared
to the period $a$. In this case the approximation of the modulation  
by the first terms of a Taylor series breaks down and its actual 
functional form becomes important. 
However, the numerical method outlined in section
\ref{secdm} is still valid provided that $s<1$, i.~e.~the total 
magnetic field  $B(x)$ is
nowhere vanishing. We consider first the noninteracting system.

\subsection{Noninteracting Electrons}
\label{secms-nonint}
The quantity we are most interested in is the
amplitude $\Delta\epsilon_{n_F}$ of the 
Landau level $n_F$ at the Fermi energy. In Fig.\ref{fig3} 
(a) (solid line) this bandwidth is shown for a weak
magnetic modulation $B_1 = 0.01 \mbox{ T}$ and average fields $B_0$ 
in the range $10\mbox{ T}> B_0 > 0.125 \mbox{ T}$. 
Starting from high fields at 
$B_0 = 10  \mbox{ T}$ we have first $n_F = 0$ 
and the bandwidth is $s\hbar\omega_0$.
When the field is lowered, $|\Delta\epsilon_{n_F}|$ increases in steps of
$2s\hbar\omega_0$ at even-integer filling factors, that is when $n_F$  
jumps by one, as follows from (\ref{aplb}). 
When $2R_c/a$ becomes larger than about $1/4$ the increase of 
$|\Delta\epsilon_{n_F}|$ becomes visibly slower 
and goes over into an oscillatory behaviour with the first 
maximum at about $2R_c =  0.6\,a$. This can be understood with 
the result (\ref{enlpt1}) of first-order perturbation theory. Using the
asymptotic relation between Laguerre polynomials and Bessel functions we
obtain for the bandwidth at the Fermi level from (\ref{enlpt1}) 
the formula \cite{Ger96a} 
\begin{equation}
  \label{denfpt}
  |\Delta\epsilon_{n_F}| \approx 
 \bigg|2\,s\,\hbar\omega_0\,A_m\,J_1(KR_c)\bigg|
\end{equation}
where $J_1$ is the Bessel function of order $1$, $R_c=l_0\sqrt{2n_F+1}$,  
and $A_m = (R_c/Kl_0^2)$.
The expression (\ref{denfpt}) describes well the bandwidths (also for small
 filling factors) as long as the average field is strong enough to ensure 
$s \ll 1$. It has zeros at approximately 
 \begin{equation}
   \label{fbcm}
   KR_c = (\lambda + 1/4)\pi \;,\,\lambda = 1,2,\ldots
 \end{equation}
corresponding to a flat-band with vanishing dispersion at the 
Fermi level.~\cite{VaPe90}  
For our parameters we encounter only the first ($\lambda=1$) of 
these magnetic flat-band situations around 
$B_0^{-1} = 6.2 \mbox{ T}^{-1}$. 
From Eq.(\ref{denfpt}) we infer further that the
maximum values of the  bandwidth at the Fermi level are of order
$2s\hbar\omega_0A_m$ rather than $2s\hbar\omega_0$.  
If we replace in $R_c$ the discrete $2n_F+1$ by $\nu$, 
the factor $A_m$ becomes
$\sqrt{a^2\rho_0/\pi g_s}$ and depends thus only on the period and 
mean density; typically we have $A_m > 10$, e.~g.~for our parameters 
$A_m = 15.6$. Consequently a seemingly weak modulation strength 
$s \sim (1/A_m)\ll 1$ is sufficient to yield for
$KR_c$ around a maximum of the Bessel function $J_1$ a bandwidth 
$|\Delta\epsilon_{n_F}| > \hbar\omega_0$ which means that the bands 
around the Fermi level do overlap. In Fig. \ref{fig4} (a)  the bandwidth 
at the Fermi level for a modulation of $B_1 = 0.1 \mbox{ T}$ is  plotted; 
it is larger than $\hbar\omega_0$ for $B_0 < 1 \mbox{ T}$. For this 
stronger modulation the spectra show, at fields $B_0 < 0.25 \mbox{ T}\,$, 
also substantial deviations from the first-order perturbation expression 
(\ref{enlpt1}), because the modulation strength $s$ 
then becomes too large. The bands around the Fermi level are not 
cosine-shaped in this regime but have
extrema away from $KX_0 = 0,\pi$. As a consequence, the
bandwidth does not go through a zero at the 
flat-band condition (\ref{fbcm}) although its behaviour still 
resembles the oscillations described by (\ref{denfpt}). 

For an electric modulation the result corresponding to (\ref{denfpt}) 
is $|\Delta\epsilon_{n_F}| = \Big| 2v_1\hbar\omega_0 J_0(KR_c)\Big|$  
with zeros at $KR_c = (\lambda - 1/4)\pi\;,\,\lambda = 1,2,\ldots$ 
(these are the electric flat-band situations) 
and the bandwidth is always smaller than $2v_1\hbar\omega_0$. In 
Fig.(\ref{fig3}) (a) the bandwidth at the Fermi level for an electric 
modulation of amplitude 
$V_1 = 0.27 \mbox{ meV} = A_m\,(\hbar e/m)\cdot 0.01 \mbox{ T}$ 
is shown as dashed line; the modulation amplitude is chosen such 
that the bandwidths are comparable to the ones induced 
by the magnetic modulation also depicted in this figure. 

In describing the induced density fluctuations 
$\delta\rho(x)=\rho(x)-\rho_0$ we 
face the difficulty that these have in general no simple shape (see
Fig.~\ref{fig1}). As a measure of their magnitude we therefore 
concentrate on their amplitude $|\Delta\rho|$. 
This quantity is displayed in units of $(1/2\pi l_0^2)$ for the 
noninteracting case at temperatures $T = 1 \mbox{ K}$ and 
$T = 0.1\mbox{ K}$ in Fig.~\ref{fig3} (b)
for a weak magnetic modulation not leading to band overlap and in
Fig.~\ref{fig4} (b) for a stronger modulation with overlapping bands. 
The lower temperature $T= 0.1 \mbox{ K}$  corresponds to 
$k_BT = 8.6\cdot 10^{-3}\mbox{  meV}$ which in the displayed range of 
$B_0$ can be considered as small
compared to $\hbar\omega_0$ whereas for the higher temperature
$T = 1 \mbox{ K}$ the finite size of $k_BT$ becomes important for about 
$B_0^{-1} > 3 \mbox{ T}^{ -1}$.  The results for the density 
can be summarized as follows: 
If the bands around the Fermi level do  not overlap and $k_BT$ is small 
against $\hbar\omega_0$ and also against the  
gap between the bands $n_F$ and $n_F+1$, then for even-integer 
filling factor the density is cosine-shaped 
$\delta\rho(x)|_{\nu\mbox{\tiny~ even}} = sr_m(KR_c)\cos(Kx)$. 
This is due to the distortion of the occupied wavefunctions by the 
modulated magnetic field.  
The amplitude $r_m$ is for $KR_c\to0$ equal to the total density $\rho_0$
(see Eq.~(\ref{dennev})) and shows for lower fields oscillations in $KR_c$
  with zeros at both electric and magnetic flat-band situations. 
During the filling of each band, i.~e.~when the filling factor 
is not an even integer, an additional density fluctuation of the order
of $(1/2\pi l_0^2)$ is produced due to the dispersion of the bands, like in
Fig.~\ref{fig1}. Both effects have a tendency to cancel each other. 
If $k_BT$ is not small compared 
to $\hbar\omega_0$ or to the gap between the bands $n_F$ and $n_F+1$,  
this cancellation becomes almost perfect and the resulting 
density modulation is minute. If the bands around the Fermi level 
do overlap (as in Fig.~\ref{fig4} (b) for $B_0 < 1 \mbox{ T}$), 
the density has a complicated shape with several extrema and its 
amplitude shows an irregular dependence on the 
filling factor. 
For $k_BT \ll \hbar\omega_0$ the amplitude of the density fluctuations 
is  still of the order of $1/2\pi l_0^2$ (but not 
larger) whereas for a higher temperature again no appreciable 
modulation of the density is produced. 

We see thus that the modulated magnetic field affects the density only 
if $R_c$ is small compared to the modulation period 
or if the temperature is very low; 
in any case the resulting density modulation is limited in amplitude 
by $g_s/2\pi l_0^2$. For not too low temperatures the 
Thomas-Fermi-prediction, 
namely no modulation of the density, holds to good accuracy 
as soon as $R_c > a/4$.  

The density amplitudes resulting from an electric modulation are shown in
Fig.~\ref{fig3} (c). As discussed in section \ref{seclpni}, 
we find for $R_c<a/8$ 
density fluctutations of order $g_s/2\pi l_0^2$ if the level $n_F$ is 
partially occupied, and essentially no modulation of the density at 
even-integer filling factors. For lower fields $B_0$, however, 
the density modulation becomes dominated by a cosine contribution whose
amplitude is not related to  $g_s/2\pi l_0^2$ but rather equals the 
Thomas-Fermi value 
\begin{equation}
  \label{delrtf}
  \delta\rho_{\mbox{\tiny TF}}(x) = - (V(x)/\mu)\rho_0 \quad.
\end{equation}
Since $\delta\rho_{\mbox{\tiny TF}}$ is independent of $B_0$ it 
appears in the plotted quantity $2\pi l_0^2 |\Delta\rho|$ as a linearly 
increasing background. 
For the higher temperature we find that for $R_c > a/4$ the density is
described accurately by Eq.~(\ref{delrtf}), 
i.~e.~$\delta\rho(x) = - (v_1\hbar\omega_0/\mu)\rho_0\cos(Kx)\,$. 
For the lower temperature deviations from this result appear which are of
similar magnitude as the corresponding deviations from zero for the 
magnetic modulation (except that at even-integer filling factors
 they do not vanish at the magnetic flat-band condition but 
only at the electric one). The density modulation for the 
low temperature and 
even-integer filling factor is entirely due to the distortion 
of the occupied wavefunctions and follows very well 
the formula derived by Aleiner and Glazman \cite{AlGl95} from first-order 
perturbation theory.
Most important for our purposes is the fact that in any case for $R_c>a/4$ 
there exists an appreciable modulation of the density, whose main part is a
wave-function effect and follows the electrostatic potential linearly,
independent of the spectrum at the Fermi level. Due to the linearity
in $V$, it is clear that this applies also to the density response 
to those higher 
Fourier components of a non-cosine electrostatic potential whose 
wavevectors $\eta K$ satisfy $\eta KR_c > 1\,$.

\subsection{Self-Consistent System}
\label{secms-sc}
Having discussed the density response induced by modulated magnetic and
electric fields, we now proceed with the investigation  of 
the effects of electrostatic
self-consistency for the magnetically modulated system. 
In Fig.~\ref{fig5} and \ref{fig6} (a) and (b)  the bandwidth at the 
Fermi level of the self-consistent system and 
the amplitude of the Hartree potential $|\Delta V^H|$ are shown
for the same parameters as used in Fig.~\ref{fig3} and \ref{fig4},
respectively. The bandwidth of the noninteracting system and 
$\hbar\omega_0$ are also shown for comparison. 
We can clearly distinguish the 
high-field regime (limited by $R_c < a/4$) discussed 
in section \ref{seclp} where the spectrum around the Fermi
level is dominated by the electrostatic effects: 
Instead of the monotonous increase in the noninteracting system 
the  bandwidth at the Fermi level is here of order $k_BT$ when the
filling factor is not close to an even-integer value and 
has at even-integer filling factors sharp maxima  
with a height of the order 
$\hbar\omega_0$ (this value is reached only for 
the stronger modulation in Fig.~\ref{fig6} while in Fig.~\ref{fig5}
Eq.~(\ref{pinkrit}) remains valid). The amplitude of the potential 
equals the difference of the self-consistent and noninteracting bandwidths,
reflecting the fact that electrostatic self-consistency is achieved by
adjusting the spectrum. 

For lower fields with $R_c>a/4$ the electrostatic corrections to the 
bandwidth at the Fermi level become much less pronounced and the 
validity of (\ref{denfpt}) is eventually restored well before the 
first magnetic flat-band situation. The main reason
for this is that now the Hartree potential is able to affect the
density independently of the dispersion at the Fermi level. 
Consequently, self-consistency can be achieved without changing 
$|\Delta\epsilon_{n_F}|$.
We first discuss the weaker modulation without band overlaps
(Fig.~\ref{fig5}). Here, for the higher temperature, the density modulation
produced by the periodic magnetic field is already small without
inclusion of the Hartree potential so that the 
self-consistent potential is also minute. 
But also for the lower termperature in (b) the situation changes
around $R_c = a/4$ and the bandwidths become close to the
noninteracting values, although the amplitude of the
self-consistent potential remains appreciable. Around $R_c = (3/8)a$ an 
electric flat-band condition is satisfied and the first 
Fourier component of
the Hartree-potential has no effect on $|\Delta\epsilon_{n_F}|$. 
Therefore here the change of the bandwidth at the Fermi 
level due to the Hartree-potential 
must be small but this is not reflected in the amplitudes of 
the self-consistent densites and potentials. 
For still lower fields, away from the electric flat-band situation, 
the Hartree potential yields again noticeable corrections 
to the bandwidths but the noninteracting curve remains essentially valid. 

For the stronger modulation displayed in Fig. (\ref{fig6}), 
the noninteracting bandwidth at the Fermi level is recovered 
as soon as it gets larger than
$\hbar\omega_0$. Since the bands at the Fermi level then overlap,  
the density modulation consists mainly of higher Fourier components 
with wave vectors $\eta K$, $\eta>1$. The induced
potential therefore belongs already to the regime of
validity of the linear relation (\ref{delrtf}) although we still have  
$R_c <a/8$. Therefore the Hartree potential here reduces the density
modulation in amplitude but does not much alter its shape, while the
bands around the Fermi level remain dominated by the cosine form imposed by
the periodic magnetic field.
\section{Conclusion}
We have calculated the density response of a 2DEG in a quantizing
magnetic field $B_0$ to a magnetic cosine modulation $B_1\cos(Kx)$ and 
compared it with the response to an electric cosine modulation. 
We also included
self-consistently the induced electrostatic potentials which reduce the
density fluctuations. We investigated in detail the changes in the energy
spectrum brought about by the requirement of electrostatic
self-consistency. In contrast to the case of an electric modulation, 
where the Hartree potential always tends to decrease the width 
of the Landau bands, for
a magnetic modulation the Hartree potential may either decrease or increase
the band dispersion, depending on strength and period of the 
modulation and on the filling factor.

In any case the produced density modulation depends
crucially on the temperature. If $k_BT$ is not small compared to
$\hbar\omega_0$ the behaviour of the density is quasi-classical, 
i.~e.~the periodic
magnetic field does not lead to an appreciable density modulation, 
while the density modulation induced by an electrostatic potential 
essentially follows the Thomas-Fermi formula. 
For temperatures satisfying $k_BT \ll \hbar\omega_0$ 
(which we assume in the remainder of this section)
both types of modulation lead to an appreciable non-classical 
inhomogeneity of the density and thus to a non-trivial 
electrostatic self-consistency problem. 

We were mainly interested in the effect of the Hartree-potential on the
spectrum around the chemical potential.  
We found that an important parameter 
is the ratio of the cyclotron radius to the length scale 
$a_{\scriptscriptstyle\!\delta\!\rho}$ of the density variation. 
If the bands in the vicinity of the Fermi level do not overlap we
have $a_{\scriptscriptstyle\!\delta\!\rho} \sim a\,$, whereas   
for overlapping bands the density fluctuations
consists of higher Fourier components and 
$a_{\scriptscriptstyle\!\delta\!\rho}$ is
significantly smaller than the period $a$ of the modulation. 
Concerning electrostatic effects we can clearly distinguish 
two regimes by the conditions 
$R_c \ll a_{\scriptscriptstyle\!\delta\!\rho}/4$ and
$R_c \gtrsim a_{\scriptscriptstyle\!\delta\!\rho}/4$. This defines 
for fixed total density a distinction between high and lower 
average magnetic fields.
The value of $B_0$ around which the regime changes depends for a weak
modulation only on the period while for a sufficiently 
strong modulation the transition takes place when the bands start to overlap. 

For $a_{\scriptscriptstyle\!\delta\!\rho}$ much larger than the 
cyclotron radius, i.~e~ for high enough $B_0$, 
the dispersion of the energy levels is
changed drastically by the inclusion of electrostatic self-consistency. 
As in the case of a purely electrostatic modulation, 
for which similar screening
effects are known,~\cite{WuGuGer88} the effective Landau bands
may be pinned to the Fermi level over regions comparable to the period. The
nonuniform magnetic field affects the density also in regions where the
chemical potential lies in a gap between two bands. In those regions the
density is not constant but reproduces the profile of the magnetic field,
since the latter alters the number of states in the vicinity of 
each center coordinate.
 
If the cyclotron radius is not small enough against 
$a_{\scriptscriptstyle\!\delta\!\rho}$, 
$R_c \gtrsim a_{\scriptscriptstyle\!\delta\!\rho}/4$, the
inclusion of electrostatic self-cosistency does not lead to an 
appreciable change of the
dispersion of the bands around the Fermi level. 
The behaviour of the latter in the regime where $R_c\sim a$ can 
therefore safely be calculated from the noninteracting system.  
\section{Acknowledgements}
We thank Daniela Pfannkuche for fruitful discussions. One of us (A.~M.) is
grateful to the Max-Planck-Institut f\"ur Festk\"orperforschung, Stuttgart,
for hospitality and support. This work was supported
by the German Bundesministerium f\"ur Bildung und Forschung (BMBF). 
%

\begin{figure}
\caption{\label{fig1}
Density $\rho(x)$ in units of $1/2\pi l_0^2$ for a magnetic 
modulation of amplitude $B_1 = 0.1 \mbox{ T}$ in average fields 
$B_0 = 2.5 - 1.7 \mbox{ T}$ corresponding to filling factors 
$\nu = 4 - 6$. The temperature is $T = 1 \mbox{ K}$. 
The dashed lines are for the noninteracting system; the 
results for $\nu = 4$ and $\nu = 6$ are marked with circles. 
The lines between these two show the successive filling of the $n=2$ 
level when $\nu$ is increased in steps of $1/3$. 
The solid lines display the results for the same filling factors 
after establishing electrostatic self-consistency.} 
\end{figure}

\begin{figure}
\caption{\label{fig2}
Results for a magnetic modulation of amplitude  
$B_1 = 0.1 \mbox{ T}$ at (a) $\nu = 5$ ($B_0 = 2.0 \mbox{ T}$), 
(b) $\nu = 4$ ($B_0 = 2.5 \mbox{ T}$) and 
(c) $\nu = 14.3$ ($B_0 =0.7\mbox{ T}$). 
The upper panel displays the spectra (dashed lines: noninteracting, 
solid lines: self-consistent solution at $T = 1\mbox{  K}$, 
dash-dotted lines: self-consistent solution at $T = 0.1\mbox{ K}$; 
the horizontal straight line with dots indicates 
the position of the chemical potential 
which, to the accuracy of the figure, is the same in all three cases). 
The lower panel shows the density
fluctuation in units of $1/2\pi l_0^2$ for the self-consistent situations
(the solid line is for $T = 1\mbox{  K}$ and the dash-dotted line 
for $T = 0.1\mbox{ K}$).} 
\end{figure}

\begin{figure}
\caption[f3]{\label{fig3} 
Bandwidth at the Fermi level and amplitude of the induced 
density fluctuations for the noninteracting systems under 
a weak magnetic modulation $B_1 = 0.01 \mbox{ T}$
and an electric modulation $V_1 = 0.27 \mbox{ meV}$ 
(leading to comparable maximal bandwidth at the Fermi level, 
i.~e.~$v_1 = A_m\,s$) for average fields 
$10\mbox{ T} > B_0 > 0.125 \mbox{ T}$. The data are
plotted versus inverse average field  indicated on the top $x$-axis. 
The bottom $x$-axis displays the ratio $2R_c/a\,\propto B_0^{-1}$.   
(a) shows the width of
the band at the Fermi level (solid line for the magnetic, 
dashed line for the electric modulation;
 the dash-dotted line is $\hbar\omega_0$). The  
amplitude of the density fluctuation in units of $1/2\pi l_0^2$ 
is shown in (b) for the magnetic and in (c) for the electric modulation. 
The solid lines are for $T =1\mbox{ K}$ and the dashed lines for 
$T = 0.1\mbox{  K}$ and the values for even-integer filling factors are 
marked with circles or diamonds, respectively. 
In (c) the dash-dotted line displays the
prediction of Thomas-Fermi-theory $|\Delta\rho| = 2(V_1/\mu)\rho_0\,$.} 
\end{figure}

\begin{figure}
\caption[f4]{\label{fig4}
Bandwidth at the Fermi level and amplitude of density modulation for the
noninteracting system as in Fig. \ref{fig3} 
but under a stronger magnetic modulation of amplitude 
$B_1 = 0.1\mbox{  T}$. In (a) the solid line is the 
width of the band at the  Fermi level and the
dash-dotted line is $\hbar\omega_0$. (b) displays the  
amplitude of the density; the solid line is for $T= 1 \mbox{ K}$ and
the dashed line for  $T = 0.1 \mbox{  K}$; the circles and diamonds 
mark the values for even-integer filling factors.} 
\end{figure}

\begin{figure}
\caption[f5]{\label{fig5}
Self-consistent results for a magnetic modulation of amplitude 
$B_1 = 0.01 \mbox{ T}$ for $10\mbox{ T} > B_0 > 0.18 \mbox{ T}$ 
plotted against $2R_c/a$. The upper two panels show for 
(a) $T= 1 \mbox{ K}$ and (b) for $T = 0.1 \mbox{  K}$ the 
bandwidth $|\Delta\epsilon_{n_F}|$ at the Fermi level 
(dashed lines with diamonds) and the amplitude of the self-consistent 
potential $|\Delta V^H|$ (dashed lines with circles). 
The filling factor was increased in steps of $1/3$ so that each 
symbol corresponds to a calculated value and the lines are only 
guides to the eye. The dash-dotted line in (a) and (b) is 
$\hbar\omega_0$ and the solid line displays the noninteracting bandwidth 
at the  Fermi level for comparison. In (c) the amplitude 
of the self-consistent densities are shown; here the solid line 
with circles is for  $T= 1 \mbox{ K}$ and the dashed line with diamonds 
for $T = 0.1 \mbox{  K}$.} 
\end{figure}

\begin{figure}
\caption[f6]{\label{fig6}
Self-consistent results for a magnetic modulation of amplitude
$B_1 = 0.1 \mbox{ T}$ and average fields 
$10\mbox{ T} > B_0 > 0.3 \mbox{ T}$. The other parameters and the 
meaning of the lines is the same as in Fig.~\ref{fig5}.} 
\end{figure}

\end{document}